\documentclass[twocolumn,showpacs,preprintnumbers,amsmath,amssymb]{revtex4}

\def\rh{r_{h}}
\def\ah{A_{h}}
\def\sh{S_{h}}
\def\mh{M_{h}}
\def\jh{J_{h}}
\def\qh{Q_{h}}
\def\th{T_{h}}
\def\omegah{\Omega_{h}}
\def\phih{\Phi_{h}}
\def\thetah{\Theta_{h}}
\def\betah{\beta_{h}}

\def\Lambdah{\Lambda_{h}}

\def\rc{r_{c}}
\def\ac{A_{c}}
\def\sc{S_{c}}
\def\mc{M_{c}}
\def\jc{J_{c}}
\def\qc{Q_{c}}
\def\tc{T_{c}}
\def\omegac{\Omega_{c}}
\def\phic{\Phi_{c}}
\def\thetac{\Theta_{c}}
\def\betac{\beta_{c}}

\def\Lambdac{\Lambda_{c}}

\begin{document}
\preprint{hep-th/0602269}

\title{Thermodynamics of de Sitter Black Holes: Thermal Cosmological Constant}
\author{Y. Sekiwa}
\email{sekiwa@stu.rikkyo.ne.jp}
\affiliation{Department of Physics, Rikkyo University, Tokyo 171-8501, Japan}

\begin{abstract}
We study the thermodynamic properties associated with the black hole event horizon and the cosmological horizon for black hole solutions in asymptotically de Sitter spacetimes. We examine thermodynamics of these horizons on the basis of the conserved charges according to Teitelboim's method. In particular, we have succeeded in deriving the generalized Smarr formula among thermodynamical quantities in a simple and natural way. We then show that cosmological constant must decrease when one takes into account the quantum effect. These observations have been obtained if and only if the cosmological constant plays the role of a thermodynamical state variable. We also touch upon the relation between inflation of our universe and a phase transition of black holes.
\end{abstract}

\pacs{04.70.Dy, 04.60.-m, 97.60.Lf}

\maketitle
\section{Introduction}
Over the past three decades, quantum field theory in de Sitter space has been a subject of growing interest. In the 1970s, the attention was due to the large symmetry group of de Sitter space. In 1980s, the focus was due to the role it played during inflation, accelerated expansion in the very early universe. Recent attention to de Sitter space and asymptotically de Sitter spacetimes is motivated by the following two aspects: First, recent cosmological observations are consistent with the possibility that there is a positive cosmological constant ($\Lambda >0$) in our universe. This possibility brings forth the picture, among many others, of some features closely related to black holes; the existence of cosmological event horizons. These are causal horizons which exist even in the absence of matter, namely in empty de Sitter space. These hide all the events which are inaccessible for each geodesic observer. Secondly, the success of the AdS/CFT correspondence \cite{Maldacena:1997re} has led to the intense study of the quantum gravity of de Sitter space \cite{Witten:2001kn}. The focus has been taken to obtain an analogue of the AdS/CFT correspondence in de Sitter space, \textit{i.e.} dS/CFT correspondence \cite{Strominger:2001pn,Strominger:2001gp}. It has been recently suggested that there is a dual relation between quantum gravity on de Sitter (dS) space and Euclidean conformal field theory (CFT) on a boundary of de Sitter space. Although there has been considerable success along this line, some theoretical obstacles exist \cite{Dyson:2002nt,Susskind:2002ri}. We do not further discuss these problems here, but emphasize that it is very important to study the gravitational systems with a cosmological constant for the quantum theory of gravity.

About thirty years ago, Hawking discovered that black holes can emit particles according to the Planck spectrum with the temperature $\kappa /2\pi$ \cite{Hawking:1974rv,Hawking:1974sw}, where $\kappa$ is the surface gravity of the black hole \cite{Bardeen:1973gs}. This means that black holes have physical temperature, not merely a quantity playing a role mathematically analogous to surface gravity in the law of black hole mechanics. The original derivation of this Hawking effect was done by making direct use of the formalism for calculating particle creation (\textit{i.e.} quantum field theory) in curved spacetime. His calculation revealed that at late times expectation number of particles at infinity is not zero, and corresponds to emission from a perfect black body at temperature $\kappa /2\pi$. Although his calculation left the microscopic process of particle creation unresolved, it became the starting point of the study based on thermodynamics for the laws of black hole mechanics. After that, many gravitational systems have been elaborated in the framework of thermodynamics, and various thermodynamic relations for black holes have been derived \cite{Wald:1995yp}.

When the gravitational systems are investigated, the important problem arises. It is whether or not the cosmological constant is a fixed parameter (or universal constant). The approach to treat the cosmological constant as a variable had already been done by Henneaux and Teitelboim many years ago \cite{Henneaux:1985tv}. They have shown that it is possible to induce cosmological constant from an antisymmetric three form gauge field coupled to the gravitational field. When the equations for the gauge field are satisfied, a cosmological constant appears as a constant of integration in the equations of motion of the coupled system. Thus the theory with the antisymmetric gauge field and without the cosmological constant is equivalent to the Einstein gravity theory with an arbitrary cosmological constant and without an antisymmetric gauge field. Henneaux and Teitelboim have shown explicitly this fact concerning anti-de Sitter spacetimes.

Recently, some authors claimed that one should regard cosmological constant $\Lambda$ as a thermodynamical variable parameter \cite{Shuang:2006eb}. They say that it is possible to consider the cosmological constant $\Lambda=\pm (D-1)(D-2)/2l^2$ as a variable parameter and promote it to a thermodynamic state variable, and that differential and integral mass formulas can be modified to
\begin{equation}
dM=TdS+\Omega dJ+\Theta dl \label{eq-intro1}
\end{equation}
and
\begin{equation}
\frac{D-3}{D-2}M=TS+\Omega J+\frac{1}{D-2}\Theta l \; , \label{eq-intro2}
\end{equation}
where $D$ is the spacetime dimensions and $\Theta$ is the generalized force conjugate to the state parameter $l$. Other authors also comment that one can regard cosmological constant $\Lambda$ as a thermodynamical variable \cite{Caldarelli:1999xj}. Although the above formulas express mathematical relations between $\Lambda$ and other thermodynamic parameters, the physical meaning of $\Lambda$ as a thermodynamical variable remains unclear.

In the present paper, we study the thermodynamical properties of black hole solutions in asymptotically de Sitter spacetimes. In particular we investigate thermodynamical law and mass formulas of these spacetimes where we treat cosmological constant $\Lambda$ as a thermodynamical state variable. Then we examine its physical meaning in addition to how cosmological constant $\Lambda$ changes. As mentioned above, the cosmological constant may be exactly explained by introducing an antisymmetric three form gauge field. Here we do not consider this microscopic behavior of the cosmological constant, but focus on the macroscopic or semiclassical behavior. So our discussions are restricted to a thermodynamical one. Most of our results are equal or similar to those in anti-de Sitter case \cite{Caldarelli:1999xj}, but the interpretation of the first law of thermodynamics is peculiar to asymptotically de Sitter spacetimes.

The organization of this paper is as follows. In section \ref{section-charges}, we consider the conserved charges for black holes in asymptotically de Sitter spacetimes. We view the black hole horizon and the cosmological horizon as two separated systems following the Euclidean black hole method in de Sitter geometry \cite{Teitelboim:2002cv,Gomberoff:2003ea} (which is closely related to the horizon thermodynamics), because these spacetimes are not in thermal equilibrium states. In general, Hawking temperatures associated with the black hole event horizon and cosmological horizon, respectively, are not equal \cite{Gibbons:1977mu}. Therefore the spacetimes for black holes in asymptotically de Sitter space will be unstable quantum mechanically. When dealing with the thermodynamics of one of two horizons, one should view the other as a boundary. Then one can obtain the conserved charges to discuss thermodynamics of two horizon spacetime. In this paper, we do not consider the details of its derivation, but give the results only. See Ref.\cite{Gomberoff:2003ea} for its calculation. In section \ref{section-thermodynamics}, we consider four-dimensional Kerr-Newman de Sitter black hole spacetime. We study thermodynamical properties associated with black hole event horizon and cosmological horizon separately using the results given in section \ref{section-charges}. Although Teitelboim's method and the derived charges are different from those used in Refs.\cite{Cai:2001sn,Cai:2001tv,Jing:2002aq,Setare:2002ss,Setare:2003jc}, it is indeed highly effectual for the Smarr formula which plays the role of the consistency condition among thermodynamical quantities. We treat cosmological constant $\Lambda$ as a thermodynamical state variable and show that integral mass formula (\ref{eq-intro2}) holds if and only if one treats the cosmological constant as a thermodynamic variable. We then show that the cosmological constant must decrease. Finally, principal conclusions and brief discussions of our results are presented in the last section.

Throughout this paper, the metric signature adopted is $(-,+,+,+)$. The use is made of natural units, namely $\hslash=c=G=1$ as well as $k=1$.
\section{Conserved Quantities}\label{section-charges}
It has been known for a long time that there exist certain difficulties when one has two or more sets of horizons with different surface gravities \cite{Gibbons:1977mu}. In our case, one must introduce separate Kruskal-like coordinate patches to cover black hole and cosmological horizons. In general, one cannot analytically continue these coordinate patches because the imaginary time periods required to avoid conical singularities at both horizons do not match. This is physically interpreted as indicating that two horizons are not in thermal equilibrium. From the point of view of the action principle, this fact means that the field equations are not satisfied everywhere. If one arranges the imaginary time period to avoid the conical singularity at the black hole horizon, the field equations will be satisfied there but will not be satisfied at the cosmological horizon. Conversely, if the role of horizons is interchanged, the field equations will not be satisfied at the black hole horizon. For black holes in asymptotically flat spacetimes, one may fix the parameters in the metric at spatial infinity, for example, mass $M$ in the case of Schwarzschild black holes. However, in the case of black holes in asymptotically de Sitter spacetimes, there is no notion of spatial infinity for observers inside the cosmological horizon, since spatial infinity exists beyond the cosmological horizon. Even if spatial infinity is accessible for the observer, he/she can not avoid conical singularities of black hole horizon and cosmological horizon at the same time. 

Following the Euclidean black hole method constructed by Teitelboim in de Sitter geometry \cite{Teitelboim:2002cv,Gomberoff:2003ea}, we view black hole horizon and cosmological horizon as two thermodynamical systems. When one discusses either one of two horizons as thermodynamical object, then the other should be viewed as a boundary. If one chooses the cosmological horizon as the boundary, where the parameters are fixed and there will be no field equations to satisfy at that point, then the problems one solves are reduced to thermodynamics of the black hole horizon contained in a space of a given cosmological horizon, which plays the analogous role of spatial infinity in the case of black holes in asymptotically flat spacetimes. Conversely, if one chooses the black hole horizon as the boundary, which plays the analogous role of coordinate origin of empty de Sitter space, then one discusses thermodynamics of cosmological horizon.

In this section, we give only the results for the conserved charges in asymptotically de Sitter spacetimes. When one discusses thermodynamics of the black hole event horizon, one must treat the cosmological horizon as a boundary. Then, for Kerr de Sitter spacetime, energy and angular momentum are given as \cite{Gomberoff:2003ea}
\begin{equation}
\mh=\frac{m}{\Xi^2} \; , \quad \jh=\frac{ma}{\Xi^2} \quad \left(\Xi=1+\frac{a^2}{l^2}\right) \; , \label{eq-charge1}
\end{equation}
where subscript ``$h$" means that it is the physical quantity associated with the black hole event horizon. These expressions have the same content as the standard Arnowitt-Deser-Misner (ADM) surface integrals for asymptotically flat spacetimes or their generalization to asymptotically anti-de Sitter spacetimes \cite{Regge:1974zd}. When the roles of the horizons are reversed, so one regards the black hole horizon as a boundary and discusses thermodynamics of cosmological horizon, the resulting expressions of energy and angular momentum change its sign;
\begin{equation}
\mc=-\frac{m}{\Xi^2} \; , \quad \jc=-\frac{ma}{\Xi^2} \; . \label{eq-charge2}
\end{equation}

One should notice the form of energy and angular momentum. First, eq.(\ref{eq-charge1}) has the same form for the case of asymptotically anti-de Sitter spacetimes if one replaces $l^2\to -l^2$ \cite{Caldarelli:1999xj}. This fact indicates that we can analytically continue the conserved charges from AdS to dS, or conversely from dS to AdS. In the latter section we see that all the results agree with those for the case of anti-de Sitter spacetimes if one replaces $l^2\to -l^2$, with respect to thermodynamic quantities for the black hole event horizon. Second, the conserved charges for the cosmological horizon are different from those for the black hole horizon only its signs. From these two facts, we can deduce that the electric charge should take the form for the black hole case and the cosmological case, respectively, as
\begin{equation}
\qh=\frac{q}{\Xi} \; , \quad \qc=-\frac{q}{\Xi} \; . \label{eq-charge3}
\end{equation}
By the same account, we take for the cosmological constant as
\begin{equation}
\Lambdah=\Lambda \; , \quad \Lambdac=-\Lambda \label{eq-charge4}
\end{equation}
where $\Lambda$ is the parameter in the metric and $\Lambdah$ and $\Lambdac$ are the physical cosmological constants which we consider as thermodynamic variables. Though the authors of Ref.\cite{Shuang:2006eb} use $l$ as a thermodynamic state variable, we use $\Lambdah$ and $\Lambdac$ as thermodynamic ones. Similarly, we use $\thetah$ and $\thetac$ as the conjugate variables to $\Lambdah$ and $\Lambdac$. In the next section, we use from eqs.(\ref{eq-charge1}) to (\ref{eq-charge4}) in order to study the thermodynamics of each horizons. For nonrotational black holes in asymptotically de Sitter spacetime, one finds $\Xi=1$, and consequently, the relations between the parameters and the physical quantities are trivial apart from its signs. For rotating case, however, these relations are very complicated.

Finally, we remark that there are the other methods to calculate the conserved quantities in asymptotically de Sitter spacetimes. For example in Refs.\cite{Cai:2001sn,Cai:2001tv,Jing:2002aq,Setare:2002ss,Setare:2003jc} the authors use the Balasubramanian-Boer-Minic (BBM) prescription \cite{Balasubramanian:2001nb} to calculate the conserved quantities for cosmological horizons and the Abbott-Deser (AD) prescription \cite{Abbott:1981ff} for black hole horizons, respectively. The BBM prescription is the method to calculate the conserved charges from stress energy tensor on the boundary. In this method, one adds the counterterm to the action in order to make the total action finite, then calculates the stress energy tensor, and last subtracts anomalous Casimir energy from gravitational mass in the case of odd spacetime dimensions. On the other hand, the AD prescription is the method to calculate by means of the deviation of metric from empty de Sitter space. The mass obtained by this method reduces to ADM mass when $\Lambda\to 0$. The conserved charges derived from these methods correspond to those derived from Teitelboim's method if the normalizations are changed. As we show in the latter sections, the thermodynamical relations are satisfied if and only if the normalizations are changed. Furthermore, for nonrotating case Teitelboim's charges are in full agreement with BBM/AD charges. So it is in the rotating case that one can decide which charges are appropriate for thermodynamics. Thus we use the Teitelboim's method, \textit{i.e.} the Euclidean black hole method in de Sitter geometry, and apply the resulting charges to the rotating black hole systems.
\section{Thermodynamic Properties}\label{section-thermodynamics}
In this section we consider the four-dimensional Kerr-Newman black hole in asymptotically de Sitter spacetime. This is the most general black hole solution of Kerr family with a positive cosmological constant. Although the generalization to higher dimensional solutions is not difficult, it is important to get the physical image by considering the four-dimensional case. The Kerr-Newman de Sitter metric can be expressed in the Boyer-Lindquist type coordinates as follows:
\begin{align}
ds^2=&-\frac{\Delta_{r}}{R^2}\left(dt-\frac{a}{\Xi}\sin^2\theta d\phi\right)^2+R^2\left(\frac{dr^2}{\Delta_{r}}+\frac{d\theta^2}{\Delta_{\theta}}\right) \nonumber \\
&\quad \quad +\frac{\Delta_{\theta}\sin^2\theta}{R^2}\left(a dt-\frac{r^2+a^2}{\Xi}d\phi\right)^2 \; , \label{eq-KN1}
\end{align}
where
\begin{align}
&R^2=r^2+a^2\cos^2\theta \; , \quad\Xi=1+\frac{a^2}{l^2} \; , \label{eq-KN2}\\
&\Delta_{r}=\left(r^2+a^2\right)\left(1-\frac{r^2}{l^2}\right)-2mr+q^2 \; , \label{eq-KN3}\\
&\Delta_{\theta}=1+\frac{a^2}{l^2}\cos^2\theta \; , \quad \frac{1}{l^2}=\frac{\Lambda}{3} \; .
\end{align}
Here $m$, $a$ and $q$ denote the mass, rotational and electric charge parameters, respectively. $\Lambda$ is the cosmological constant parameter. We treat $\Lambda$ as a variable parameter. The metric (\ref{eq-KN1}) solves the Einstein-Maxwell field equations with electromagnetic vector given by
\begin{equation}
A_t=\frac{q r}{R^2} \; , \quad A_{\phi}=\frac{q r}{R^2\Xi}a\sin^2\theta \; . \label{eq-KN4}
\end{equation}
For $q=0$, $a=0$, both $q=0$ and $a=0$, this metric evidently reduces to the Kerr de Sitter metric, the Reissner-Nordstr$\ddot{{\rm o}}$m de Sitter metric, the Schwarzschild de Sitter metric, respectively. The horizons of the Kerr-Newman de Sitter spacetime follow from the equation $\Delta_r=0$. This algebraic equation has the four roots which are three positive and one negative solutions in the condition that the relation
\begin{align}
&\left[(l^2-a^2)^2-12l^2(a^2+q^2)\right]^3 > \nonumber \\
&\; \left[(l^2-a^2)^3+36l^2(l^2-a^2)(a^2+q^2)-54m^2l^4\right]^2 \label{eq-KN5}
\end{align}
is satisfied for the parameters $m$, $a$, $q$ and $l$. The largest positive solution is the cosmological horizon $\rc$, the smallest positive solution is inner black hole horizon (\textit{i.e.} inner Cauchy horizon), and the other positive solution is the black hole event horizon $\rh$. The negative solution has no physical meaning. In this paper, we assume that eq.(\ref{eq-KN5}) is satisfied, so the metric (\ref{eq-KN1}) represents Kerr-Newman black hole in asymptotically de Sitter spacetime. Since we do not study the internal structure of Kerr-Newman de Sitter black holes, the horizons we are interested in here are black hole horizon $\rh$ and cosmological horizon $\rc$. In the following we investigate thermodynamic properties of the black hole event horizon and the cosmological horizon separately.
\subsection{Black hole event horizon}\label{section-thermodynamics1}
First, we discuss thermodynamics of the black hole event horizon $\rh$. Then we must treat the cosmological horizon $\rc$ as a boundary where the parameters are fixed. The case for which the role of horizons is reversed is discussed in the next subsection.

The physical mass $\mh$, the physical angular momentum $\jh$, the physical electric charge $\qh$ and the physical cosmological constant $\Lambdah$ are related to the parameters $m$, $a$, $q$ and $\Lambda$ from eqs.(\ref{eq-charge1}) to (\ref{eq-charge4}) as follows:
\begin{equation}
m=\Xi^2\mh \; , \quad a=\frac{\jh}{\mh} \; , \quad q=\Xi\qh \; , \quad \Lambda=\Lambdah \; . \label{eq-bh-charge}
\end{equation}
The area of the black hole horizon is written as
\begin{equation}
\ah=\int\left|g_{\theta\theta}g_{\phi\phi}\right|_{r=\rh}^{1/2}d\theta d\phi =\frac{4\pi(\rh^2+a^2)}{\Xi} \; . \label{eq-bh-area}
\end{equation}
Analytical continuation of the Lorentzian metric by $t\to -i\tau$ and $a\to ia$ yields the Euclidean section \cite{Hawking:1998kw}, whose regularity at $r=\rh$ requires that we must identify $\tau\sim\tau +\betah$ and $\phi\sim\phi+i\betah\omegah'$. This postulate of Euclidean regularity determines the inverse Hawking temperature $\betah$ and the angular velocity $\omegah'$ of the black hole horizon as follows:
\begin{align}
\betah&=-\frac{4\pi l^2\rh(\rh^2+a^2)}{3\rh^4+(a^2-l^2)\rh^2+l^2(a^2+q^2)} \; , \label{eq-bh-beta} \\
\omegah'&=\frac{a\,\Xi}{\rh^2+a^2} \; . \label{eq-bh-omega'}
\end{align}
The Bekenstein-Hawking entropy $\sh$ is associated with $\betah$ (or $\th$) through
\begin{equation}
\betah=\left(\frac{\partial\sh}{\partial\mh}\right)_{\jh\qh\Lambdah} \; \bigg[\;\th=\left(\frac{\partial \mh}{\partial \sh}\right)_{\jh\qh\Lambdah}\bigg] \label{eq-bh-S/M}
\end{equation}
at constant angular momentum $\jh$, electric charge $\qh$ and cosmological constant $\Lambdah$, which yields
\begin{equation}
\sh=\frac{\pi(\rh^2+a^2)}{\Xi} \label{eq-bh-S}
\end{equation}
for entropy. Thus the so-called Bekenstein-Hawking relation between entropy and area of the horizon, \textit{i.e.} $\sh=\ah/4$ holds \cite{Bekenstein:1973ur}.

The angular velocity of the black hole horizon is eq.(\ref{eq-bh-omega'}) in the present coordinates. However, this is not appropriate for thermodynamics. The angular velocity $\,\omegah\,$ relevant to the Kerr-Newman de Sitter black hole thermodynamics is indeed defined $\grave{\textit{a}}$ \textit{la} Christodoulou and Bekenstein as follows \cite{Smarr:1972kt}:
\begin{equation}
\omegah=\left(\frac{\partial \mh}{\partial \jh}\right)_{\sh\qh\Lambdah}=-\th\left(\frac{\partial \sh}{\partial \jh}\right)_{\mh\qh\Lambdah} \label{eq-bh-omega1}
\end{equation}
at constant mass $\mh$, electric charge $\qh$ and cosmological constant $\Lambdah$. Then we get the thermal angular velocity of black hole event horizon as
\begin{equation}
\omegah=\frac{a}{\rh^2+a^2}\left(1-\frac{\rh^2}{l^2}\right) \; . \label{eq-bh-omega2}
\end{equation}
This angular velocity has an extra rotation compared with $\omegah'$ in eq.(\ref{eq-bh-omega'}). It is not the relative angular velocity $\omegah'-\omegac'$ of the black hole horizon $\rh$ relative to the cosmological horizon $\rc$, as one might have naively expected, where $\omegac'$ is given by eq.(\ref{eq-ch-omega'}). Since eq.(\ref{eq-bh-omega2}) is written as
\begin{equation}
\omegah=\frac{a\,\Xi}{\rh^2+a^2}\;-\;\frac{a}{l^2} \; , 
\end{equation}
we find that eq.(\ref{eq-bh-omega2}) is the angular velocity of the rotating black hole relative to $r=\infty$, because the term $-a/l^2$ is what one obtains if one sets $r=\infty$ in $N^{\phi}$ (for the definition and detailed expression of $N^{\phi}$, see Ref.\cite{Gomberoff:2003ea}). Of course $r=\infty$ is not in the Euclidean section, but eq.(\ref{eq-bh-omega2}) is precisely the analytical continuation of the case of anti-de Sitter spacetimes \cite{Caldarelli:1999xj}. This indicates that it is possible to analytically continue the angular velocity from anti-de Sitter case to de Sitter case in similar way for the conserved charges.

The electric potential $\phih$ is also defined $\grave{\textit{a}}$ \textit{la} Christo\-doulou and Bekenstein as follows:
\begin{equation}
\phih=\left(\frac{\partial \mh}{\partial \qh}\right)_{\sh\jh\Lambdah}=-\th\left(\frac{\partial \sh}{\partial \qh}\right)_{\mh\jh\Lambdah} \label{eq-bh-phi1}
\end{equation}
at constant mass $\mh$, angular momentum $\jh$ and cosmological constant $\Lambdah$. Then we get the electrical potential of black hole event horizon as
\begin{equation}
\phih=\frac{\rh q}{\rh^2+a^2} \; . \label{eq-bh-phi2}
\end{equation}
This is consistent with $A_t=qr/R^2$ which is the solution of the Maxwell equation in the Kerr-Newman de Sitter spacetimes. $A_t$ equals to $\phih$ at the black hole horizon.

In similar way, the variable $\thetah$ conjugate to cosmological constant $\Lambdah$ is defined as
\begin{equation}
\thetah=\left(\frac{\partial \mh}{\partial \Lambdah}\right)_{\sh\jh\qh}=-\th\left(\frac{\partial \sh}{\partial \Lambdah}\right)_{\mh\jh\qh} \label{eq-bh-theta1}
\end{equation}
at constant mass $\mh$, angular momentum $\jh$ and electric charge $\qh$. Then we get
\begin{equation}
\thetah=-\frac{1}{6l^2\Xi^2}\left[ma^2l^2+\rh(\rh^2+a^2)(l^2+a^2)\right] \; . \label{eq-bh-theta2}
\end{equation}
To examine the meaning of this thermodynamic quantity, we take nonrotational limit $a\to 0$. Then $\thetah$ is written as $\thetah=-\rh^3/6$. In this limit, spacetime is spherically symmetric so that this corresponds to the volume of region which is occupied by black hole, except to prefactor. To show this, we consider the combination $\thetah\Lambdah$. As the quantity $\thetah$ is conjugate to cosmological constant $\Lambdah$, this combination term has dimension of energy. Since cosmological constant has vacuum energy density $\Lambdah /8\pi$, it is reasonable that $\thetah$ has dimension of volume. If we rewrite $\thetah\Lambdah$ as $-(4\pi\rh^3/3)(\Lambdah /8\pi)$, this corresponds to the product between the vacuum energy density and the volume occupied by black hole. In this sense, we interpret eq.(\ref{eq-bh-theta2}) as the volume inside the event horizon of the black hole for the rotating case, except to prefactor. Thus we call $\thetah$ (or $\thetac$) the generalized volume with respect to the black hole (cosmological) horizon.

As shown by Henneaux and Teitelboim \cite{Henneaux:1985tv}, if the cosmological constant is expressed by a three form gauge field, the variable conjugate to the three form gauge field should be a three form. Accordingly, one can expect that the quantity $\thetah$ is related to this conjugate three form. Since, in this paper, we consider only the macroscopic or semiclassical behavior of the cosmological constant as a thermodynamical variable, we do not investigate the microscopic behavior of these three forms, any further.

Using eqs.(\ref{eq-bh-charge}) and (\ref{eq-bh-S}), one can obtain a simple mass formula of the black hole event horizon as
\begin{equation}
\mh^2=\left(\frac{\pi}{\sh}-\frac{1}{l^2}\right)\jh^2+\frac{\sh}{4\pi}\left(\frac{\pi\qh^2}{\sh}+1-\frac{\sh}{\pi l^2}\right)^2 . \label{eq-bh-smarr}
\end{equation}
This is the generalized Smarr formula of the black hole event horizon. We can then claim that the present formalism automatically satisfies the consistency condition $\grave{\textit{a}}$ \textit{la} the Smarr formula among natural thermodynamical quantities. It contains as usual all the information about the thermodynamic state of black hole. If we use the AD prescription to calculate the conserved quantities \cite{Abbott:1981ff}, by the way, we can not obtain the generalized Smarr formula of this form. Thus we can say that the Teitelboim's method is consistent with thermodynamics. Note that the above generalized Smarr formula has the same form for anti-de Sitter case \cite{Caldarelli:1999xj}. If one replaces $l^2\to -l^2$, eq.(\ref{eq-bh-smarr}) agrees with the generalized Smarr formula for AdS case completely.  This fact suggests that we can analytically continue from AdS to dS, or conversely, from dS to AdS, and that at least thermodynamically there are some relations between anti-de Sitter spacetimes and de Sitter spacetimes. We expect that these relations and its physical meaning will be revealed by the quantum theory of gravity.

If we regard $\mh$ as a function of $\sh$, $\jh$, $\qh^2$ and $\Lambdah^{-1}$, it is a homogeneous function of degree $1/2$. Applying Euler's theorem we obtain
\begin{equation}
\frac{1}{2}\mh=\th\sh+\omegah\jh+\frac{1}{2}\phih\qh-\thetah\Lambdah \; . \label{eq-bh-euler}
\end{equation}
As mentioned above, the other formalism does not provide us the generalized Smarr formula. If the form of the generalized Smarr formula is different from eq.(\ref{eq-bh-smarr}), the physical mass $\mh$ may not be a homogeneous function of degree $1/2$, and consequently eq.(\ref{eq-bh-euler}) is not derived. Suppose that $\Lambdah$ is not a thermodynamical variable and $\mh$ is not as a function of $\Lambdah^{-1}$. Then eq.(\ref{eq-bh-euler}) is not derived, even if the generalized Smarr formula takes the form eq.(\ref{eq-bh-smarr}). Indeed, authors of Ref.\cite{Gibbons:2004ai} did not consider the cosmological constant as a thermodynamic variable and they could not get the AdS version of eq.(\ref{eq-bh-euler}). Therefore we assert that $\Lambdah$ must be a thermodynamic variable and by eq.(\ref{eq-charge4}) cosmological constant $\Lambda$ also must be a variable parameter. This holds, however, for only the case that one treats black hole system quantum mechanically or semiclassically. In general relativity, $\Lambda$ must be constant because the Einstein tensor is divergenceless. As mentioned in introduction, eq.(\ref{eq-bh-euler}) is an integral mass formula [see eq.(\ref{eq-intro2})]. In four dimension, eqs.(\ref{eq-intro2}) and (\ref{eq-bh-euler}) agree for the uncharged case (note that we use $\Lambdah$ as a thermodynamic state variable). Therefore, we find that at least semiclassically cosmological constant $\Lambda$ is not constant, but a variable parameter.

One can define the quantities conjugate to $\sh$, $\jh$, $\qh$ and $\Lambdah$ from the generalized Smarr formula of black hole event horizon. These are the temperature
\begin{align}
\th&=\frac{1}{8\pi \mh}\bigg[1-\frac{\pi^2}{\sh^2}\left(4\jh^2+\qh^4\right) \nonumber \\
&\qquad \qquad \qquad -\frac{2}{l^2}\left(\qh^2+\frac{2\sh}{\pi}\right)+\frac{3\sh^2}{\pi^2 l^4}\bigg] \; , \label{eq-bh-beta3}
\end{align}
the angular velocity
\begin{equation}
\omegah=\frac{\pi \jh}{\mh\sh}\left(1-\frac{\sh}{\pi l^2}\right) \; , \label{eq-bh-omega3}
\end{equation}
the electric potential
\begin{equation}
\phih=\frac{\pi \qh}{2\mh\sh}\left(\qh^2+\frac{\sh}{\pi}-\frac{\sh^2}{\pi^2 l^2}\right) \label{eq-bh-phi3}
\end{equation}
and the generalized volume
\begin{equation}
\thetah=-\frac{1}{2\mh}\left[\frac{1}{3}\jh^2+\frac{\sh}{6\pi}\left(\qh^2+\frac{\sh}{\pi}\right)-\frac{\sh^3}{6\pi^3 l^2}\right]  , \label{eq-bh-theta3}
\end{equation}
respectively. If we use, instead of $\sh$, $\jh$, $\qh$ and $\Lambdah$, horizon radius and parameters in the metric, eqs.(\ref{eq-bh-beta3}) to (\ref{eq-bh-theta3}) correspond to, of course, eqs.(\ref{eq-bh-beta}), (\ref{eq-bh-omega2}), (\ref{eq-bh-phi2}) and (\ref{eq-bh-theta2}). All these thermodynamic quantities are similar to those of the anti-de Sitter case \cite{Caldarelli:1999xj}. If one replaces $l^2$ to $-l^2$ in the above thermodynamic quantities, one can get the AdS version of these thermodynamic quantities.

Let us turn our attention to the first law of thermodynamics. From eqs.(\ref{eq-bh-S/M}), (\ref{eq-bh-omega1}), (\ref{eq-bh-phi1}) and (\ref{eq-bh-theta1}) the first law of thermodynamics for the black hole event horizon is expressed as follows:
\begin{equation}
d\mh={\th}d\sh+{\omegah}d\jh+{\phih}d\qh+{\thetah}d\Lambdah \, . \label{eq-bh-first}
\end{equation}
This law means that total energy of the black hole system is conserved.

We shall first consider the case of the classical process version, where ``classical" means that its physics is allowed to be described by general relativity only. By the Bekenstein-Hawking entropy-area law, the first law of thermodynamics is written as follows:
\begin{equation}
d\mh+{\omegah}d(-\jh)+{\phih}d(-\qh)=\frac{\kappa_h}{8\pi}d\ah \; , \label{eq-bh-first-c}
\end{equation}
where the term $\thetah d\Lambdah$ is ignored because the cosmological constant is treated as a fixed constant in general relativity. Eq.(\ref{eq-bh-first-c}) implies that energy variation follows from the classical area increasing law for the black hole event horizon. The first term of left hand side contributes to the area increasing as the black hole mass increasing. The second and third terms contribute as extraction of rotational and electric potential energies. That is to say, when one extracts rotational or electric energy from the black hole, its area increases by eq.(\ref{eq-bh-first-c}), which is well-known as the Penrose process in asymptotically flat spacetimes \cite{Penrose:1969pc}.

Next, we consider the case of the quantum process version, where ``quantum" means that we take Hawking effect into consideration. For the sake of simplicity, we consider the uncharged and nonrotating case only. Then the first law of thermodynamics is written as
\begin{equation}
-d\mh+\frac{4\pi\rh^3}{3}\,d\left(-\frac{\Lambdah}{8\pi}\right)=-\th d\sh \; , \label{eq-bh-first-q}
\end{equation}
where we have changed the signs of both sides. The left hand side expresses the mass loss of black hole and the decrease of vacuum energy inside the black hole event horizon (This is because black hole horizon radius $\rh$ shrinks when black hole mass or cosmological constant decreases). Then eq.(\ref{eq-bh-first-q}) means that, for the observer outside the black hole event horizon, the total energy variation is seen as the decrease of entropy inside the black hole event horizon. This energy (or entropy) is carried away from inside to outside by means of Hawking radiation (its temperature is $\th$). Though the decrease of entropy contradicts the second law of thermodynamics, by the generalized second law, the possibility of this quantum process is sustained. Now we consider the effect and its physical meaning of the second term of left hand side in eq.(\ref{eq-bh-first-q}). We assume $\mh$ is fixed. Then eq.(\ref{eq-bh-first-q}) suggests that the decrease of vacuum energy density is equal to the entropy decreasing of the black hole event horizon. Because the black hole horizon radius $\rh$ shrinks when cosmological constant decreases, this phenomenon can be seen by the outside observer as if black hole radiates its energy and consequently generalized entropy increases. Since the generalized second law of thermodynamics requires that the generalized entropy increases for the all physical processes, the cosmological constant must decrease. Therefore we can conclude that vacuum energy is transformed quantum mechanically to the energy of radiation by means of decaying cosmological constant. These are the phenomena which does not happen classically, \textit{i.e.} in general relativity.
\subsection{Cosmological horizon}\label{section-thermodynamics2}
In the previous subsection, we have studied the thermodynamic properties associated with the black hole event horizon $\rh$. In this subsection, we discuss thermodynamics of the cosmological horizon $\rc$ along the similar line. Here one must reverse the role of horizons. So we treat the black hole event horizon as a boundary where the parameters are fixed. The calculation is almost the same with the one in the previous case, but mathematically, the signs of some equations and horizon radius are changed ($\rh \to \rc$). The physical meaning of the first law of thermodynamics is a little modified.

From eqs.(\ref{eq-charge1}) to (\ref{eq-charge4}), the physical mass $\mc$, the physical angular momentum $\jc$, the physical electric charge $\qc$ and the physical cosmological constant $\Lambdac$ are related to the parameters $m$, $a$, $q$ and $\Lambda$ as follows:
\begin{equation}
m=-\Xi^2\mc \; , \;\, a=\frac{\jc}{\mc} \; , \;\, q=-\Xi\qc \; , \:\, \Lambda=-\Lambdac \; . \label{eq-ch-charge}
\end{equation}
The area of the cosmological horizon is written as
\begin{equation}
\ac=\int\left|g_{\theta\theta}g_{\phi\phi}\right|_{r=\rc}^{1/2}d\theta d\phi =\frac{4\pi(\rc^2+a^2)}{\Xi} \; .
\end{equation}
Analytical continuation of the Lorentzian metric by $t\to -i\tau$ and $a\to ia$ yields the Euclidean section, whose regularity at $r=\rc$ requires that we must identify $\tau\sim\tau +\betac$ and $\phi\sim\phi+i\betac\omegac'$. This postulate of Euclidean regularity determines the inverse Hawking temperature $\betac$ and the angular velocity $\omegac'$ of the cosmological horizon $\rc$ as follows:
\begin{align}
\betac&=\frac{4\pi l^2\rc(\rc^2+a^2)}{3\rc^4+(a^2-l^2)\rc^2+l^2(a^2+q^2)} \; ,\label{eq-ch-beta} \\
\omegac'&=\frac{a\,\Xi}{\rc^2+a^2} \; .\label{eq-ch-omega'}
\end{align}
The Bekenstein-Hawking entropy $\sc$ is associated with $\betac$ (or $\tc$) through
\begin{equation}
\betac=\left(\frac{\partial\sc}{\partial\mc}\right)_{\jc\qc\Lambdac} \; \bigg[\;\tc=\left(\frac{\partial \mc}{\partial \sc}\right)_{\jc\qc\Lambdac}\bigg] \label{eq-ch-S/M}
\end{equation}
at constant angular momentum $\jc$, electric charge $\qc$ and cosmological constant $\Lambdac$, which yields
\begin{equation}
\sc=\frac{\pi(\rc^2+a^2)}{\Xi} \label{eq-ch-S}
\end{equation}
for entropy. Thus the so-called Bekenstein-Hawking relation between entropy and area of the horizon, \textit{i.e.} $\sc=\ac/4$ holds also for the cosmological horizon.

The angular velocity of the cosmological horizon is eq.(\ref{eq-ch-omega'}) in the present coordinates. However this is not appropriate for thermodynamics. In the same way for the black hole case, the angular velocity $\omegac$ relevant to thermodynamics is defined $\grave{\textit{a}}$ \textit{la} Christodoulou and Bekenstein as follows:
\begin{equation}
\omegac=\left(\frac{\partial \mc}{\partial \jc}\right)_{\sc\qc\Lambdac}=\frac{a}{\rc^2+a^2}\left(1-\frac{\rc^2}{l^2}\right) \; . \label{eq-ch-omega1}
\end{equation}
This thermal angular velocity is not the angular velocity of cosmological horizon relative to the black hole horizon. $\omegac$ is an angular velocity relative to the coordinate origin inside the black hole horizon because eq.(\ref{eq-ch-omega1}) is written as
\begin{equation}
\omegac=\frac{a\,\Xi}{\rc^2+a^2}\;-\;\frac{a}{l^2} \; ,
\end{equation}
where the second term equals to $N^{\phi}$ at $r=0$ (see Ref.\cite{Gomberoff:2003ea}). In this case, however, there exist no obvious interpretation in terms of an analytic continuation from anti-de Sitter spacetimes since there does not exist a cosmological horizon in asymptotically anti-de Sitter spacetimes.

The electric potential $\phic$ is also defined $\grave{\textit{a}}$ \textit{la} Christo\-doulou and Bekenstein as follows:
\begin{equation}
\phic=\left(\frac{\partial \mc}{\partial \qc}\right)_{\sc\jc\Lambdac}=\frac{\rc q}{\rc^2+a^2} \; . \label{eq-ch-phi1}
\end{equation}
This is consistent with $A_t=qr/R^2$ which is the solution of the Maxwell equation in the Kerr-Newman de Sitter spacetimes. $A_t$ equals to $\phic$ at the cosmological horizon.

In similar way, the generalized volume $\thetac$ is defined as
\begin{align}
\thetac&=\left(\frac{\partial \mc}{\partial \Lambdac}\right)_{\sc\jc\qc} \nonumber \\
&=-\frac{1}{6l^2\Xi^2}\left[ma^2l^2+\rc(\rc^2+a^2)(l^2+a^2)\right] \; . \label{eq-ch-theta1}
\end{align}
This corresponds to the volume inside the cosmological horizon in the same sense of eq.(\ref{eq-bh-theta2}).

Using eqs.(\ref{eq-ch-charge}) and (\ref{eq-ch-S}), one can obtain a simple mass formula of cosmological horizon as
\begin{equation}
\mc^2=\left(\frac{\pi}{\sc}-\frac{1}{l^2}\right)\jc^2+\frac{\sc}{4\pi}\left(\frac{\pi\qc^2}{\sc}+1-\frac{\sc}{\pi l^2}\right)^2 . \label{eq-ch-smarr}
\end{equation}
This is the generalized Smarr formula of cosmological horizon. This is the same form with the generalized Smarr formula for the case of black hole horizon. This suggests that the present formalism automatically satisfies the consistency condition $\grave{\textit{a}}$ \textit{la} the Smarr formula among natural thermodynamical quantities in the same way as the black hole case. If we use the BBM prescription to calculate the conserved quantities \cite{Balasubramanian:2001nb}, we can not obtain the generalized Smarr formula of this form. Thus we can say that the Teitelboim's method is consistent with thermodynamics for the cosmological horizon also. Note that, for the anti-de Sitter case, there exists no analogue of the above generalized Smarr formula. It is not obvious what the Smarr formula means when one replaces $l^2\to -l^2$ in the above equation.

If we regard $\mc$ as a function of $\sc$, $\jc$, $\qc^2$ and $\Lambdac^{-1}$, it is a homogeneous function of degree $1/2$. Applying Euler's theorem we obtain
\begin{equation}
\frac{1}{2}\mc=\tc\sc+\omegac\jc+\frac{1}{2}\phic\qc-\thetac\Lambdac \; . \label{eq-ch-euler}
\end{equation}
If $\Lambdac$ is not thermodynamical variable and $\mc$ is not as a function of $\Lambdac^{-1}$, eq.(\ref{eq-ch-euler}) is not derived. Therefore we assert that $\Lambdac$ must be a thermodynamic variable and by eq.(\ref{eq-charge4}) cosmological constant $\Lambda$ also must be a variable parameter. These facts are the same for the case of the black hole event horizon. Therefore, from the viewpoint of thermodynamics for both horizons, it is concluded that cosmological constant must be a variable parameter.

One can define the quantities conjugate to $\sc$, $\jc$, $\qc$ and $\Lambdac$ from the generalized Smarr formula of cosmological horizon. These are the temperature
\begin{align}
\tc=\frac{1}{8\pi \mc}&\bigg[1-\frac{\pi^2}{\sc^2}\left(4\jc^2+\qc^4\right) \nonumber \\
& \quad -\frac{2}{l^2}\left(\qc^2+\frac{2\sc}{\pi}\right)+\frac{3\sc^2}{\pi^2 l^4}\bigg] \; , \label{eq-ch-beta3}
\end{align}
the angular velocity
\begin{equation}
\omegac=\frac{\pi \jc}{\mc\sc}\left(1-\frac{\sc}{\pi l^2}\right) \; , \label{eq-ch-omega3}
\end{equation}
the electric potential
\begin{equation}
\phic=\frac{\pi \qc}{2\mc\sc}\left(\qc^2+\frac{\sc}{\pi}-\frac{\sc^2}{\pi^2 l^2}\right) \label{eq-ch-phi3}
\end{equation}
and the generalized volume
\begin{equation}
\thetac=-\frac{1}{2\mc}\left[\frac{1}{3}\jc^2+\frac{\sc}{6\pi}\left(\qc^2+\frac{\sc}{\pi}\right)-\frac{\sc^3}{6\pi^3 l^2}\right] \, , \label{eq-ch-theta3}
\end{equation}
respectively. These formulas for the cosmological horizon are similar to those for the black hole case.

Let us turn our attention to the first law of thermodynamics. From eqs.(\ref{eq-ch-S/M}), (\ref{eq-ch-omega1}), (\ref{eq-ch-phi1}) and (\ref{eq-ch-theta1}) the first law of thermodynamics for the cosmological horizon is expressed as follows:
\begin{equation}
d\mc={\tc}d\sc+{\omegac}d\jc+{\phic}d\qc+{\thetac}d\Lambdac \, . \label{eq-ch-first}
\end{equation}
This law means that the total energy inside the cosmological horizon is conserved.

First, for simplicity, we consider the uncharged and nonrotating case in similar way to the case of black hole event horizon. The first law is then written as
\begin{equation}
d\mc={\tc}d\sc+{\thetac}d\Lambdac \; . \label{eq-ch-sdsfirst}
\end{equation}
Furthermore, if one specializes to the case $\mc=0$ or $m=0$, the first law becomes
\begin{equation}
\frac{4\pi\rc^3}{3}\,d\left(-\frac{\Lambda}{8\pi}\right)=\tc d\sc \; . \label{eq-ch-ds}
\end{equation}
This is the first law of thermodynamics for the cosmological horizon in empty de Sitter space. One should notice that the left had side expresses the increase of vacuum energy inside the cosmological horizon, and this is the increase of vacuum energy that observer can see. Indeed, $4\pi\rc^3/3$ is the volume of the visible region for observer. One may note that for the black hole case entropy expresses the information which the observer cannot see, \textit{i.e.} the information in the region inside black hole. If we suppose that the analogue of black hole case holds for the cosmological case, we deduce that entropy $\sc$ for the cosmological horizon should express the information which observer cannot see, \textit{i.e.} the information in the region outside the cosmological horizon ($\rc < r$). Eq.(\ref{eq-ch-ds}) indicates that the entropy increase of the cosmological horizon results from the energy increase of the visible region ($0 < r < \rc$). This is different from the black hole case because the energy increase inside the cosmological horizon contributes to the entropy increase. But the information outside the cosmological horizon do not contribute. One can deduce, however, that at the outside of cosmological horizon the energy density decreases, because the cosmological constant is independent of spacetime coordinates. If $\Lambda$ decreases inside the cosmological horizon, it should decrease at the outside also. Although the detailed mechanism for the decrease of vacuum energy still remains mysterious, we can expect that the cosmological constant must decrease in order to hold the second law of thermodynamics which say entropy $\sc$ never decreases. From eq.(\ref{eq-ch-ds}), one finds that entropy increases if and only if cosmological constant $\Lambda$ decreases. Therefore cosmological constant $\Lambda$ decreases through the quantum mechanical effect. We attribute the origin of Hawking radiation from the cosmological horizon to the decay of the cosmological constant, at least semiclassically.

One may note that the quantum process version considered above is similar to the classical one for the cosmological horizon, because entropy (horizon area) decreasing process is forbidden by the generalized second law of thermodynamics (by the area theorem). In general relativity, the cosmological constant must be constant. Thus the area increasing process does not happen for empty de Sitter space. So we have no need to consider the classical process version.

Next, we return to the case $\mc\neq 0$. If a little modification is added to eq.(\ref{eq-ch-sdsfirst}), the first law is written as follows:
\begin{align}
&\bigg[\frac{4\pi\rc^3}{3}\,d\left(\frac{\Lambdac}{8\pi}\right)-\frac{4\pi\rh^3}{3}\,d\left(-\frac{\Lambdah}{8\pi}\right)\bigg] \nonumber \\
& \qquad +\bigg[\frac{4\pi\rh^3}{3}\,d\left(-\frac{\Lambdah}{8\pi}\right)-d\mh\bigg]=\tc d\sc \; . \label{eq-ch-SdSfirst1}
\end{align}
The first square bracket of the left hand side corresponds to the increase of vacuum energy inside the cosmological horizon but outside the black hole horizon (because $\rc$ increases and $\rh$ shrinks when $\Lambda$ decreases, total vacuum energy in the region $\rh < r < \rc$ increases). The second square bracket expresses the radiational energy from black hole in the sense of eq.(\ref{eq-bh-first-q}). Although in the Euclidean black hole geometry method when one discusses the black hole horizon as a boundary, the region which corresponds to the inside of black hole horizon is removed from the manifold, thermodynamical law holds if and only if one considers as if the removed region exists. If the second bracket is written by the entropy of the black hole event horizon, the first law of thermodynamics takes the form where the physical meaning is more explicit,
\begin{align}
\frac{4\pi\rc^3}{3}\,d\left(-\frac{\Lambda}{8\pi}\right)&-\frac{4\pi\rh^3}{3}\,d\left(-\frac{\Lambda}{8\pi}\right) \nonumber \\
&\quad \qquad -\th d\sh=\tc d\sc \; . \label{eq-ch-SdSfirst2}
\end{align}
This equation implies that increase of the entropy for the cosmological horizon is due to the thermal radiation from two event horizons, \textit{i.e.} black hole horizon and cosmological horizon. The first and second terms of left hand side are the increase of vacuum energy in the region $\rh <r<\rc$. On the other hand, $-\th d\sh$ represents the entropy loss inside the black hole horizon. That is to say, the origins of entropy increase are the increase of vacuum energy in the visible region and Hawking radiation from black hole. As a whole, entropy must increase by the generalized second law of thermodynamics. Eq.(\ref{eq-ch-SdSfirst2}) indicates these phenomena explicitly. Therefore we find that $\sc$ expresses the generalized entropy in the visible region, and that $\sc$ increases if and only if $\Lambda$ decreases. We can say again that $\Lambda$ must decrease through the quantum effect.

It is straightforward to include the effects of electric charge and angular momentum. If electric charge is taken into account, the first law of thermodynamics becomes as follows
\begin{align}
&\bigg[\frac{4\pi\rc^3}{3}\,d\left(-\frac{\Lambda}{8\pi}\right)-\frac{4\pi\rh^3}{3}\,d\left(-\frac{\Lambda}{8\pi}\right)\bigg] \nonumber \\
&\quad \qquad +\bigg(\frac{q}{\rc}-\frac{q}{\rh}\bigg)dq-\th d\sh=\tc d\sc \; . \label{eq-ch-RNdSfirst}
\end{align}
The first square bracket of the left hand side corresponds to the increase of vacuum energy. The second bracket expresses the electric potential difference between the cosmological horizon and the black hole horizon. This term implies that electric energy is extracted from the inside to the outside of the black hole horizon, because it is interpreted, by the factor $dq$, as the increase of electric energy in the visible region. $-\th d\sh$ represents the entropy loss of black hole. In addition to the electric charge, when the effect of angular momentum is included, the first law is written as
\begin{align}
&\left(\thetac-\thetah\right)d(-\Lambdac)+\left(\omegac-\omegah\right)d(-\jc) \nonumber \\
& \qquad +\left(\phic-\phih\right)d(-\qc)-\th d\sh=\tc d\sc \; .
\end{align}
The first term of the left hand side corresponds to the increase of vacuum energy. Here $\thetac$ and $\thetah$ have the rotational effects. The second term expresses the extracted rotational energy. The angular velocity is the one of the black hole horizon relative to the cosmological horizon. Similarly, the third term represents the extracted electric energy. Finally, the last term is due to Hawking radiation from black hole. These express explicitly that both the energy decrease inside the black hole and the energy increase in the visible region contribute to the increase of entropy for the cosmological horizon. Thus it is confirmed that $\sc$ expresses the generalized entropy in the visible region $(\rh <r<\rc)$.
\section{Conclusion and Discussion}\label{section4}
In the present paper, we have studied the thermodynamic properties associated with the black hole event horizon and the cosmological horizon for black hole solutions in asymptotically de Sitter spacetimes. Principal results are as follows.

First of all, it must be emphasized that we have considered the black hole horizon and the cosmological horizon as two thermodynamical systems. We have then found that each horizon can be treated as a thermodynamical object in spite of the fact that black hole solutions in asymptotically de Sitter spacetimes are not in thermodynamical equilibrium as a whole system. We have made use of the Euclidean black hole method in de Sitter geometry according to Teitelboim \cite{Teitelboim:2002cv,Gomberoff:2003ea} to calculate the conserved quantities. One of the features of the conserved quantities in asymptotically de Sitter spacetimes is that the parameters in the metric and the physical (conserved) quantities are complicatedly related to each other for the case of rotating black holes in the same way as for the case of black holes in anti-de Sitter spacetimes. The other feature is that these conserved charges correspond to those for the case of anti-de Sitter spacetimes, if one replaces $l^2$ to $-l^2$. This fact indicates that we may be able to analytically continue from AdS to dS, or from dS to AdS. There are certain differences for the cosmological horizons, however. Since black hole solutions in asymptotically anti-de Sitter spacetimes have no cosmological horizons, one cannot discuss the conserved quantities of cosmological horizons in the similar fashion as for asymptotically de Sitter spacetimes.

Second, we have studied thermodynamics of black hole and cosmological horizons separately. The results obtained from these considerations indicate that one can discuss black hole horizons and cosmological horizons of black hole solutions in asymptotically de Sitter spacetimes on the basis of thermodynamics. The macroscopic entropy-area law $S=A/4$ which relates thermodynamic entropy to the area of event horizon is universally valid for any types of black holes belonging to the Kerr family. Understanding the microscopic origin of this law is undoubtedly a key step towards understanding the fundamental nature of spacetime. We have then built up a set of natural thermodynamical quantities. These thermodynamical quantities (temperature, entropy, angular velocity and electric potential) are similar to those for the case of black holes in anti-de Sitter spacetimes with respect to the black hole horizons. If one replaces $l^2$ to $-l^2$ for the black hole cases, these quantities correspond to those for the case of black holes in anti-de Sitter spacetimes \cite{Caldarelli:1999xj}. Again, it may be possible to analytically continue from AdS to dS, or conversely from dS to AdS with respect to the black hole horizons. Furthermore, we have succeeded in establishing the generalized Smarr formula for the mass as a function of entropy, angular momentum, electric charge and cosmological constant in the sense of the consistency condition among these natural thermodynamical quantities. This fact implies that thermodynamical quantities mentioned above surely satisfy the first and second laws of thermodynamics associated with the black hole horizons and cosmological horizons, respectively. The generalized Smarr formula for the black hole event horizon in asymptotically de Sitter spacetimes has a similar form to the one in asymptotically anti-de Sitter spacetimes. Indeed, the generalized Smarr formula derived in this paper can be obtained by replacing $l^2$ to $-l^2$ for the one of the AdS case \cite{Caldarelli:1999xj}. Thermodynamic relations of both horizons are really assured to be consistent if and only if the cosmological constant is considered as a variable parameter. If one treats cosmological constant as a fixed constant, one cannot obtain an integral mass formula. Suppose one uses the BBM/AD prescriptions, instead of Teitelboim's method, to calculate the conserved quantities. Even if one considers the cosmological constant as a variable, then, the full set of natural thermodynamical quantities are not obtained and consequently the generalized Smarr formula is not assured.

Finally, we have investigated the first law of thermodynamics not only for the black hole horizon but also for the cosmological horizon. We have revealed that the cosmological constant must decrease if quantum mechanical effect is taken into account. The decrease of the cosmological constant explains the increase of vacuum energy in the region which the observer can see. We find that this is the energy content of radiation from the cosmological horizon, and that this is consistent with the generalized second law of thermodynamics. In other words, thermodynamic laws are valid if and only if the cosmological constant decreases. When the cosmological constant decreases, the energy increases inside the cosmological horizon, on the other hand, the energy outside the cosmological horizon decreases. This is because energy density decreases everywhere. The detailed mechanism of the energy decreasing outside the cosmological horizon still remains unresolved, however. We can only claim that this results from the decrease (or the decay) of the cosmological constant. Our result yields an antipodal viewpoint against the conventional dS/CFT correspondence which claims quantum gravity in de Sitter space with fixed cosmological constant in the sense of dual representation as a conformally invariant Euclidean field theory on the boundary of de Sitter space \cite{Strominger:2001pn,Strominger:2001gp}. Since even semiclassical theory considering the background spacetimes as classical geometry makes the cosmological constant decrease, it is questionable that quantum gravity with a fixed cosmological constant can be established.

In the precedent paragraphs, we have summarized the most important prospects of our thermodynamical investigation of black hole event horizons and cosmological horizons. Let us now briefly touch upon the further thermodynamical aspects beyond the scope of the present investigation. Hawking temperatures associated with the black hole event horizon and cosmological horizon are not equal each other, in general. Therefore, the spacetime for black holes in asymptotically de Sitter space is not in thermal equilibrium. It will be possible to expect that thermal equilibrium is eventually brought to realization in the future. Thus we will be led to suspect that the phase transition of black holes arises in the similar fashion as for the anti-de Sitter spacetimes \cite{Hawking:1982dh} in which black holes evaporate into a hot gas, or equivalently, event horizons of black holes disappear at critical temperature through the so-called Hawking-Page phase transition. We are now investigating whether or not the Hawking-Page like phase transition is really materialized in de Sitter black hole spacetimes, in general. Recently, Carlip and Vaidya \cite{Carlip:2003ne} have afforded qualitative confirmation to the realization of the Hawking-Page like phase transition in Reissner-Nordstr$\ddot{\textrm{o}}$m de Sitter spacetimes. The detailed phase structure of black hole in asymptotically de Sitter spacetimes is left totally unresolved, however. Next, one may remember that black hole horizons and cosmological horizons thermodynamically resemble each other. We can then expect that if black holes in asymptotically de Sitter spacetimes undergo the Hawking\--Page like phase transition, the cosmological horizons will also undergo something like the Hawking\--Page phase transition. If cosmological horizons undergo the Hawking-Page like phase transition, then, the cosmological horizons will disappear. This phenomenon is the phase transition of vacuum and means that exponential expansion (inflation) through which one can not see far distant region will stop. Any observer can see spatial infinity, in principle, after this phase transition. These tantalizing enigmata are highly expected to be resolved in the future investigation.
\begin{acknowledgments}
I would like to thank Professor H. Fujisaki for helpful discussions and valuable suggestions.
\end{acknowledgments}

\end{document}